\documentclass[prd,twocolumn, amsmath,amssymb,floatfix,nofootinbib]{revtex4}
\usepackage{mathrsfs,bm}
\usepackage{longtable,lscape}
\usepackage{txfonts}
\usepackage{indentfirst}
\usepackage{graphicx,color,dcolumn,booktabs}
\usepackage{multirow}
\usepackage{indentfirst}
\usepackage{multirow}
\usepackage{epsfig}
\usepackage{graphicx,color,dcolumn}
\usepackage{epstopdf}
\usepackage{amsmath}
\usepackage{diagbox}
\usepackage{changes}
\usepackage{threeparttable}
\usepackage{booktabs}
\usepackage{color}
\usepackage{amssymb}
\definecolor{cover}{rgb}{0.77,0.87,0.88}
\definecolor{blueone}{rgb}{0.1,0.1,.7}
\definecolor{citec}{rgb}{0.14,0.47,0.09}
\definecolor{two}{rgb}{0.0,0.5,0.}
\definecolor{three}{rgb}{.5,.1,0.15}
\usepackage[bookmarks=true,bookmarksopen=false,plainpages=false,breaklinks=true,
   bookmarksnumbered=true,hypertexnames=false,
   filecolor=blue,urlcolor=three,menucolor=three,
   linkcolor=three,citecolor=blueone, colorlinks,
   anchorcolor=blue,runcolor=pink,frenchlinks=red
   pdfstartview=FitH,pdftitle=title,%
   pdfauthor=author]{hyperref}

\usepackage{relsize}
\usepackage{xspace}
\def\babar{\mbox{\slshape B\kern-0.1em{\smaller A}\kern-0.1em
    B\kern-0.1em{\smaller A\kern-0.2em R}}}

\newcolumntype{C}{>{$}c<{$}}
\AtBeginDocument{
\heavyrulewidth=.08em
\lightrulewidth=.05em
\cmidrulewidth=.03em
\belowrulesep=.65ex
\belowbottomsep=0pt
\aboverulesep=.4ex
\abovetopsep=0pt
\cmidrulesep=\doublerulesep
\cmidrulekern=.5em
\defaultaddspace=.5em
}

\allowdisplaybreaks[4]

\begin{document}
\title{Possible charmed-strange molecular dibaryons}
\author{Shu-Yi Kong$^1$, Jun-Tao Zhu$^1$, Jun He$^{1,2}$\footnote{Corresponding author: junhe@njnu.edu.cn}}

\affiliation{$^1$School of Physics and Technology, Nanjing Normal University, Nanjing 210097, China\\
$^2$Lanzhou Center for Theoretical Physics, Lanzhou University, Lanzhou 730000, China}

\date{\today}
\begin{abstract}

In this work, we systematically investigate the dibaryons  with charm number
$C$=1 and strangeness number $S$=$\pm$ 1 from the interactions of a charmed
baryon and a strange baryon $\Lambda_c\Lambda$, $\Lambda_c\Sigma^{(*)}$,
$\Sigma_c^{(*)}\Lambda$, and $\Sigma^{(*)}_c\Sigma^{(*)}$, and corresponding
interactions of a charmed baryon and an antistrange baryon
$\Lambda_c\bar{\Lambda}$, $\Lambda_c\bar{\Sigma}^{(*)}$,
$\Sigma^{(*)}_c\bar{\Lambda}$, and $\Sigma^{(*)}_c\bar{\Sigma}^{(*)}$. With the
help of the effective Lagrangians with $SU(3)$, heavy quark, and chiral
symmetries, the potentials of the interactions considered are constructed by
light meson exchanges. To search for the possible molecules, the quasipotential
Bethe-Salpeter equation with the interaction potential kernel is solved to find
poles from scattering amplitude. The results suggest that  attractions widely
exist in charmed-strange system with $C$=1 and $S$=$-1$. The $S$-wave bound
states can be produced from most of the channels.  Few bound states are also
produced from the charmed-antistrange interactions. Couple-channel effect are
considered in the current work to discuss the couplings of the molecular states
to the channels considered. More experimental research for these charmed-strange
dibaryons are suggested.

\end{abstract}

\maketitle

\section{INTRODUCTION}

In the last two decades, a growing number of exotic particles have been
discovered in experiment. It is hard to put these exotic particles into the
conventional quark model and their inner structures are  still under debate.  It
implies other interpretations such as molecular states, compact multiquarks,
hybrids and some nonresonant state interpretations.  Inspired by the observation
that many exotic particles were observed near the thresholds of two
hadrons, the molecular state picture, which is
a shallow bound state of two or more hadrons, naturally becomes a popular
interpretation of the exotic particles.

All the time, it is a big challenge to find possible dibaryon molecules, which
is an important part of the spectrum of the hadronic molecular states.  The
well-known deuteron can be seen as a  dibaryon molecular state composed of two
nucleons. Jaffe suggested another famous dibaryon, $H$ dibaryon, which is a
bound state of the $\Lambda\Lambda$ system with $[uuddss]$
configuration~\cite{Jaffe:1976yi}.  Actually, the idea of possible dibaryon
molecules can go back to 1964. Dyson and Xuong predicted dibaryon states based
on the SU(6) symmetry ~\cite{Dyson:1964xwa}. In their prediction, the mass of
dibaryon $\Delta\Delta(D_{03})$ was 2376~MeV, and the mass of dibaryon
$N\Delta(D_{21}, D_{21})$ was 2176~MeV. The dibaryon $D_{03}$ predicted has a
mass surprisingly close to the later discovery of $d^*(2380)$ at
WASA~\cite{WASA-at-COSY:2011bjg}. The resonant structure was studied in many
theoretical
methods~\cite{Gal:2014zia,Gal:2013dca,Huang:2014kja,Haidenbauer:2011za,Park:2015nha,Dong:2015cxa,Dong:2016rva},
including studies of assigning it as a molecular state from the $\Delta\Delta$
interaction~\cite{Haidenbauer:2011za}. However, such an assumption leads to a
binding energy of about 80~MeV, which tends to assign it as a compact hexaquark
rather than a bound state of two $\Delta$ baryons. The peak structure was
suggested to be a triangle singularity in the last step of the
reaction~\cite{Ikeno:2021frl,Molina:2021bwp}, which can be traced back to early
work in Ref.~\cite{Bar-Nir:1973mxc}. The WASA-at-COSY Collaboration reported a
hint of an isotensor dibaryon with quantum numbers $IJ^P=2(1^+)$ with a mass of
2140~MeV~\cite{WASA-at-COSY:2018zlh}, which is slightly below the $N\Delta$
threshold. In our previous work~\cite{Lu:2020qme}, we studied the possible
S-wave molecular states from the $N\Delta$ interaction within the quasipotential
Bethe-Salpeter equation (qBSE) approach. The results suggest a $D_{21}$ bound
state can be produced from the interaction and there may exist two more possible
$D_{12}$ and $D_{22}$ states with smaller binding energies. More theoretical and
experimental works are required to clarify existence of these states and the
origin of the experimentally observed states. 

There also exist some studies about the molecular state with a baryon and an
antibaryon. Such molecular states can carry quantum numbers as a meson.  For
example, the $X(2239)$~\cite{BESIII:2018ldc} and $\eta(2225)$~\cite{DM2:1986wyp}
have  mass of 2239~MeV and 2220~MeV, respectively, which are very close to  a
pair of baryon $\Lambda$ and  antibaryon $\bar{\Lambda}$. Hence, they are good
candidates for  hidden-strange baryon $\Lambda\bar{\Lambda}$ molecular states.
In Ref.~\cite{Zhao:2013ffn,Zhu:2019ibc}, two  bound states with spin parities
$1^-$ and $0^-$ from the $\Lambda\bar{\Lambda}$ interaction were assigned to the
$X(2239)$ and the $\eta(2225)$, respectively. There are also many investigations
to interpret  the $X(2239)$ and the $\eta(2225)$ in other
picutres~\cite{Lu:2019ira,Azizi:2019ecm, Li:2008we,Wang:2017iai}.

In the heavy flavor sector, the hidden-charm and double-charm dibaryons also
enter the view of the community of exotic hadrons even though there are not many
relevant experimental results right now.  It is not difficult to understand that
heavy quark systems are more likely to have attraction and form bound states in
the molecular state picture due to the reduction of the kinetic of the systems
resulting from the large masses of the heavy quark systems. Up to now, some
theoretical investigations have been performed to look for  possible bound
states composed of  two heavy baryons or a pair of heavy  baryon and antibaryon
within the  constituent quark
model~\cite{Carames:2015sya,Lu:2022myk,Vijande:2016nzk}, one-boson-exchange model~\cite{Meng:2017fwb, Li:2012bt,Dong:2021juy}, a quark level effective
potential~\cite{Chen:2021cfl}, a chromo-magnetic interaction
model~\cite{Liu:2021gva}. In Ref.~\cite{Song:2022svi}, we performed a systematic
study of possible molecular states composed of two charmed baryons. The results
suggest that strong attractions exist in both hidden-charm and double-charm
systems and bound states can be produced in most of the systems.  All these
theoretical results support the existence of bound states of heavy flavor
dibaryons.

The systems mentioned above are all composed of two light baryons or two heavy
baryons. Besides systems above, it is still a blank for the study of systems
composed of a light and a heavy baryon. In fact, there exist a large amount of
study about the systems with a light meson and heavy meson. For example, the
experimentally observed $D^*_{s0}(2317)$, $D_{s1}(2460)$, $X_0(2900)$ are
considered as the candidates of the $DK$
molecule~\cite{Barnes:2003dj,Oset:2016cnj,Rosner:2006vc,Zhang:2006ix,Hofmann:2003je,Kong:2021ohg},
$D^*K$ molecule~\cite{Guo:2006rp,Rosner:2006vc,Hofmann:2003je,Kong:2021ohg}, and
$\bar{D}^*K^*$
molecule~\cite{Liu:2020nil,Chen:2020aos,Agaev:2020nrc,Huang:2020ptc,Kong:2021ohg},
respectively.  Hence, it is nature to investigate the systems composed of a
charmed and a strange/antistrange baryon. In this work, we systematically
investigate the charmed-strange interactions of a charmed baryon and a strange
baryon $\Lambda_c\Lambda$, $\Lambda_c\Sigma^{(*)}$, $\Sigma_c^{(*)}\Lambda$, and
$\Sigma^{(*)}_c\Sigma^{(*)}$, and corresponding charmed-antistrange interactions
of a charmed baryon and an antistrange baryon $\Lambda_c\bar{\Lambda}$,
$\Lambda_c\bar{\Sigma}^{(*)}$, $\Sigma^{(*)}_c\bar{\Lambda}$, and
$\Sigma^{(*)}_c\bar{\Sigma}^{(*)}$ in the qBSE approach to study the
possibilities of existence of charmed-strange molecules.

The work is organized as follows. After introduction, the potential kernels of
charmed-strange baryon systems are presented, which is obtained with the help of
the effective Lagrangians with $SU(3)$, heavy quark, and chiral symmetries. And
the qBSE approach will be introduced briefly. In Section \ref{Sec:BcB}, the
results for the molecular states from the charmed-strange interactions are
presented with both single and coupled channel calculations. The Section
\ref{Sec:BcBbar}  contributes to posbbile molecular states from the
charmed-antistrange interactions. In Section \ref{Sec: summary}, discussion and
summary are given.

\section{Theoretical frame}\label{Sec: Formalism}

To study the charmed-strange systems considered in the current work and the
couplings between different channels, the potential will be constructed
within the one-boson-exchange model. The exchanges by peseudoscalar mesons
$\mathbb{P}$, vector mesons $\mathbb{V}$ and  scalar meson $\sigma$ will be
considered. Hence, the  Lagrangians  depicting the couplings of  charmed or
strange baryons with light mesons are required.

\subsection{ Relevant Lagrangians}

For the strange part, we consider the exchange of $\pi$, $\eta$, $\rho$,
$\omega$, and $\sigma$ mesons with the strange baryons $\Lambda$, $\Sigma$ and
$\Sigma^*$ respectively. In the current work, the $\phi$ exchange does not
contribute due to the suppression by the OZI rule. For the former four mesons,
the interaction can be described by the effective Lagrangians with SU(3) and
chiral symmetries~\cite{Ronchen:2012eg,Kamano:2008gr}. The explicit forms can
be written as,
\begin{eqnarray}
\mathcal{L}_{BBP}&=& -\frac{g_{BBP}}{m_P}\bar{B}\gamma^{5}\gamma^{\mu}\partial_{\mu}PB,\\
\mathcal{L}_{BBV}&=&-\bar{B}\left[g_{BBV}\gamma^{\mu}-\frac{f_{BBV}}{2m_{B}}\sigma^{\mu\nu}\partial_{\nu}\right]V_{\mu}B,\\
\mathcal{L}_{B^*B^*P}&=&\frac{g_{B^*B^*P}}{m_P}\bar{B^*}_{\mu}\gamma^{5}
\gamma^{\nu}B^{*\mu}\partial_{\nu}P, \\
\mathcal{L}_{B^*B^*V}&=&-\bar{B}^*_{\tau}\left[g_{B^*B^*V}\gamma^{\mu}-\frac{f_{B^*B^*V}}
{2m_{B^*}}\sigma^{\mu\nu}\partial_{\nu}\right]V_{\mu} B^{*\tau},\\
\mathcal{L}_{BB^*P}&=&\frac{g_{BB^*P}}{m_{P}}\bar{B}^{*\mu} \partial_{\mu}PB\;+\; \text{h.c.},\\
\mathcal{L}_{BB^*V}&=&-i \frac{g_{BB^*V}}{m_{V}}\bar{B}^{*\mu}\gamma^5\gamma^{\nu} V_{\mu\nu}B\;+\; \text{h.c.},
\end{eqnarray}
where $V_{\mu\nu}={\partial_{\mu}}\vec{V}_{\nu}-{\partial_{\mu}}\vec{V}_{\mu}$, and the coupling constants can be determined by the SU(3) symmetry~\cite{Ronchen:2012eg,deSwart:1963pdg,Lu:2020qme,Zhu:2022fyb}. The SU(3) relations and the explicit values of coupling constants are listed in Table~\ref{constants}.

 \renewcommand\tabcolsep{0.06cm}
\renewcommand{\arraystretch}{1.45}
\begin{table}
\centering
\caption{The coupling constants in effective Lagrangians. Here,
$g_{BBP}=g_{NN\pi}=0.989$, $g_{BBV}=g_{NN\rho}=3.25$, $g_{{B^*B^*}P}=\sqrt{60} g_{\Delta\Delta\pi}=13.78$, $g_{{B^*B^*}V}=\sqrt{60} g_{\Delta\Delta\rho}=59.41$, $g_{B{B^*}P}=\sqrt{20} g_{N\Delta\pi}=9.48$, $g_{B{B^*}V}=\sqrt{20} g_{N\Delta\rho}=71.69$, $\alpha_{P}=0.4$, $\alpha_{V}=1.15$, $f_{NN\rho}= g_{NN\rho}\kappa_{\rho}$, $f_{\Delta\Delta\rho}= g_{\Delta\Delta\rho}\kappa_{\rho}$ with $\kappa_{\rho}=6.1$, $f_{NN\omega}=0$~\cite{Ronchen:2012eg,Lu:2020qme,Zhu:2022fyb}.\label{constants}}
\scalebox{0.99}{
\begin{tabular}[t]{lrr||lrr}\bottomrule[1.5pt]
 Coupling 				&SU(3) Relation		& Values			 				&Coupl.			&SU(3) Relation 		& Values	 \\ \hline
 $g_{\Lambda\Lambda\omega}$ &$\frac{2}{3}(5\alpha_V-2)g_{BBV}$&$8.12$
 &$f_{\Lambda\Lambda\omega}$&$\frac{5}{6}f_{NN\omega}-\frac{1}{2}f_{NN\rho}$&$-9.9$\\
$g_{\Sigma\Sigma\pi}$&$2\alpha _P g_{BBP}$&$0.79$
& $g_{\Sigma\Sigma\eta}$&$\frac{2}{\sqrt{3}}(1-\alpha_P) g_{BBP}$&$0.68$\\
$g_{\Sigma\Sigma\rho/\omega}$&$2\alpha_V g_{BBV}$&$7.47$
&$f_{\Sigma\Sigma\rho/\omega}$&$\frac{1}{2}f_{NN\omega}+\frac{1}{2}f_{NN\rho}$&$9.9$\\
$g_{\Sigma^*\Sigma^*\pi}$&$\frac{1}{2\sqrt{15}} g_{B^*B^*P}$&$1.78$
&$g_{\Sigma^*\Sigma^*\eta}$&$0$&$0$\\
$g_{\Sigma^*\Sigma^*\rho/\omega}$&$\frac{1}{2\sqrt{15}} g_{B^*B^*V}$&$7.67$
&$f_{\Sigma^*\Sigma^*\rho/\omega}$&$\frac{1}{2\sqrt{15}}f_{\Delta\Delta\rho}$&$46.78$\\
$g_{\Lambda\Sigma\pi}$&$\frac{2}{\sqrt{3}}(1-\alpha_P) g_{BBP}$&$0.68$
&$g_{\Lambda\Sigma\rho}$&$\frac{2}{\sqrt{3}}(1-\alpha_V)g_{BBV}$&$-0.56$\\
$g_{\Lambda\Sigma^*\pi}$&$\frac{1}{2\sqrt{10}}g_{BB^*P}$&$1.49$
&$g_{\Lambda\Sigma^*\rho}$&$\frac{1}{2\sqrt{10}}g_{BB^*V}$&$11.33$\\
$g_{\Sigma\Sigma^*\pi}$&$\frac{1}{2\sqrt{30}}g_{BB^*P}$&$0.86$
&$g_{\Sigma\Sigma^*\eta}$&$\frac{1}{2\sqrt{10}}g_{BB^*P}$&$1.49$\\
$g_{\Sigma\Sigma^*\rho/\omega}$&$\frac{1}{2\sqrt{30}}g_{BB^*V}$&$6.54$
&$g_{\Sigma\Sigma^*\omega}$&$\frac{1}{2\sqrt{30}}g_{BB^*V}$&$6.54$\\\bottomrule[1.5pt]\hline
\end{tabular}
}
\end{table}

For the coupling of strange baryons with the scalar meson $\sigma$, the
Lagrangians are~\cite{Zhao:2013ffn} \begin{eqnarray}
\mathcal{L}_{BB\sigma}&=&-g_{BB\sigma}\bar{B}\sigma B,\\
\mathcal{L}_{B^*B^*\sigma}&=&g_{B^*B^*\sigma}\bar{B}^{*\mu}\sigma B^*_{\mu}.
\end{eqnarray} The different choices of the mass of $\sigma$ meson from 400 to
550 MeV affects the result a little, which can be smeared by a small variation
of the cutoff. In this work, we adopt a $\sigma$ mass of 500~MeV. In
general, we choose the coupling constants $g_{BB\sigma}$ and $g_{B^*B^*\sigma}$
as the same value as $g_{BB\sigma}=g_{B^*B^*\sigma}=6.59$~\cite{Zhao:2013ffn}.

For the charmed part, the Lagrangians for the couplings between the charmed
baryons and exchanged mesons can be constructed under the heavy quark and chiral
symmetry~\cite{Cheng:1992xi,Yan:1992gz,Wise:1992hn,Casalbuoni:1996pg}. The
explicit forms of the Lagrangians can be written as,
\begin{eqnarray}
{\cal L}_{BB\mathbb{P}}&=&-\frac{3g_1}{4f_\pi\sqrt{m_{\bar{B}}m_{B}}}~\epsilon^{\mu\nu\lambda\kappa}\partial^\nu \mathbb{P}~
\sum_{i=0,1}\bar{B}_{i\mu} \overleftrightarrow{\partial}_\kappa B_{j\lambda},\nonumber\\
{\cal L}_{BB\mathbb{V}}&=&-i\frac{\beta_S g_V}{2\sqrt{2m_{\bar{B}}m_{B}}}\mathbb{V}^\nu
 \sum_{i=0,1}\bar{B}_i^\mu \overleftrightarrow{\partial}_\nu B_{j\mu}\nonumber\\
&-&i\frac{\lambda_S
g_V}{\sqrt{2}}(\partial_\mu \mathbb{V}_\nu-\partial_\nu \mathbb{V}_\mu) \sum_{i=0,1}\bar{B}_i^\mu B_j^\nu,\nonumber\\
{\cal L}_{BB\sigma}&=&\ell_S\sigma\sum_{i=0,1}\bar{B}_i^\mu B_{j\mu},\nonumber\\
{\cal L}_{B_{\bar{3}}B_{\bar{3}}\mathbb{V}}&=&-i\frac{g_V\beta_B}{2\sqrt{2m_{\bar{B}_{\bar{3}}}m_{B_{\bar{3}}}} }\mathbb{V}^\mu\bar{B}_{\bar{3}}\overleftrightarrow{\partial}_\mu B_{\bar{3}},\nonumber\\
{\cal L}_{B_{\bar{3}}B_{\bar{3}}\sigma}&=&
\ell_B \sigma \bar{B}_{\bar{3}}B_{\bar{3}},\nonumber\\
{\cal L}_{BB_{\bar{3}}\mathbb{P}}
    &=&-i\frac{g_4}{f_\pi} \sum_i\bar{B}_i^\mu \partial_\mu \mathbb{P} B_{\bar{3}}+{\rm H.c.},\nonumber\\
{\cal L}_{BB_{\bar{3}}\mathbb{V}}    &=&\frac{g_\mathbb{V}\lambda_I}{\sqrt{2 m_{\bar{B}}m_{B_{\bar{3}}}}}\epsilon^{\mu\nu\lambda\kappa} \partial_\lambda \mathbb{V}_\kappa\sum_i\bar{B}_{i\nu} \overleftrightarrow{\partial}_\mu
   B_{\bar{3}}+{\rm H.c.},\label{LB}
\end{eqnarray}
where $S^{\mu}_{ab}$ is composed of Dirac spinor operators as,
\begin{eqnarray}
    S^{ab}_{\mu}&=&-\sqrt{\frac{1}{3}}(\gamma_{\mu}+v_{\mu})
    \gamma^{5}B^{ab}+B^{*ab}_{\mu}\equiv{ B}^{ab}_{0\mu}+B^{ab}_{1\mu},\nonumber\\
    \bar{S}^{ab}_{\mu}&=&\sqrt{\frac{1}{3}}\bar{B}^{ab}
    \gamma^{5}(\gamma_{\mu}+v_{\mu})+\bar{B}^{*ab}_{\mu}\equiv{\bar{B}}^{ab}_{0\mu}+\bar{B}^{ab}_{1\mu},
\end{eqnarray}
and the charmed baryon matrices are defined as,
\begin{align}
B_{\bar{3}}&=\left(\begin{array}{ccc}
0&\Lambda^+_c&\Xi_c^+\\
-\Lambda_c^+&0&\Xi_c^0\\
-\Xi^+_c&-\Xi_c^0&0
\end{array}\right),\quad
B=\left(\begin{array}{ccc}
\Sigma_c^{++}&\frac{1}{\sqrt{2}}\Sigma^+_c&\frac{1}{\sqrt{2}}\Xi'^+_c\\
\frac{1}{\sqrt{2}}\Sigma^+_c&\Sigma_c^0&\frac{1}{\sqrt{2}}\Xi'^0_c\\
\frac{1}{\sqrt{2}}\Xi'^+_c&\frac{1}{\sqrt{2}}\Xi'^0_c&\Omega^0_c
\end{array}\right), \nonumber\\
B^*&=\left(\begin{array}{ccc}
\Sigma_c^{*++}&\frac{1}{\sqrt{2}}\Sigma^{*+}_c&\frac{1}{\sqrt{2}}\Xi^{*+}_c\\
\frac{1}{\sqrt{2}}\Sigma^{*+}_c&\Sigma_c^{*0}&\frac{1}{\sqrt{2}}\Xi^{*0}_c\\
\frac{1}{\sqrt{2}}\Xi^{*+}_c&\frac{1}{\sqrt{2}}\Xi^{*0}_c&\Omega^{*0}_c
\end{array}\right).\label{MBB}
\end{align}
The $\mathbb P$ and $\mathbb V$ are the pseudoscalar and vector matrices as,
\begin{equation}
    {\mathbb P}=\left(\begin{array}{ccc}
        \frac{\sqrt{3}\pi^0+\eta}{\sqrt{6}}&\pi^+&K^+\\
        \pi^-&\frac{-\sqrt{3}\pi^0+\eta}{\sqrt{6}}&K^0\\
        K^-&\bar{K}^0&-\frac{2\eta}{\sqrt{6}}
\end{array}\right),
\mathbb{V}=\left(\begin{array}{ccc}
\frac{\rho^0+\omega}{\sqrt{2}}&\rho^{+}&K^{*+}\\
\rho^{-}&\frac{-\rho^{0}+\omega}{\sqrt{2}}&K^{*0}\\
K^{*-}&\bar{K}^{*0}&\phi
\end{array}\right).\nonumber \label{MPV}
\end{equation}

The coupling constants in the above Lagrangians are listed in Table~\ref{coupling}, which are cited from the literatures~\cite{Chen:2019asm, Liu:2011xc, Isola:2003fh, Falk:1992cx}.
 \renewcommand\tabcolsep{0.425cm}
\renewcommand{\arraystretch}{1.}
\begin{table}[h!]
\caption{The parameters and coupling constants. The $\lambda$, $\lambda_{S,I}$ and $f_\pi$ are in the unit of GeV$^{-1}$. Others are in the unit of $1$.
\label{coupling}}
\begin{tabular}{cccccccccccccccccc}\bottomrule[1.5pt]
$f_\pi$&$g_V$&$\beta_S$&$\ell_S$&$g_1$\\
0.132 &5.9&-1.74&6.2&-0.94\\\hline
$\lambda_S$ &$\beta_B$&$\ell_B$ &$g_4$&$\lambda_I$\\
-3.31&$-\beta_S/2$&$-\ell_S/2$&$g_1/{2\sqrt{2}\over 3}$&$-\lambda_S/\sqrt{8}$ \\
\bottomrule[1.5pt]
\end{tabular}
\end{table}

\subsection{Potential kernel of interactions}

With the above Lagrangians for the vertices, the potential kernel can be
constructed in the one-boson-exchange model with the help of the standard
Feynman rule as in Refs.~\cite{He:2019ify,He:2015mja}. The propagators of the
exchanged light mesons are defined as, \begin{eqnarray}%
P_\mathbb{P,\sigma}(q^2)&=&
\frac{i}{q^2-m_\mathbb{P,\sigma}^2}~f_i(q^2),\nonumber\\
P^{\mu\nu}_\mathbb{V}(q^2)&=&i\frac{-g^{\mu\nu}+q^\mu
q^\nu/m^2_{\mathbb{V}}}{q^2-m_\mathbb{V}^2}~f_i(q^2), \end{eqnarray} where the
form factor $f_i(q^2)$ is adopted to compensate the off-shell effect of
exchanged meson, which is in form of $e^{-(m_e^2-q^2)^2/\Lambda_e^4}$ with $m_e$
and $q$ being the mass and  momentum of the exchanged light mesons,
respectively.

In this work, we do not give the explicit form of the potential due to the large
number of channels to be considered. Instead, we input the vertices $\Gamma$ and
the above propagators $P$ into the code directly and the potential can be
constructed with the help of the standard
Feynman rule as~\cite{He:2019ify},
\begin{eqnarray}%
{\cal V}_{\mathbb{P},\sigma}=I_{\mathbb{P},\sigma}\Gamma_1\Gamma_2 P_{\mathbb{P},\sigma}(q^2),\quad
{\cal V}_{\mathbb{V}}=I_{\mathbb{V}}\Gamma_{1\mu}\Gamma_{2\nu}  P^{\mu\nu}_{\mathbb{V}}(q^2).
\end{eqnarray}
The $I_{\mathbb{P},\mathbb{V},\sigma}$ is the flavor factors of the certain
meson exchange as listed in Table~\ref{flavor factor}.
The interaction of charmed-antistrange interactions will be rewritten to the
charmed-strange interactions by the well-known G-parity
rule~\cite{Phillips:1967zza,Klempt:2002ap},
\begin{eqnarray}
V&=&\sum_{i}{\zeta_{i}V_{i}}=-V_{\pi}+V_{\eta}+V_{\rho}-V_{\omega}+V_{\sigma}.\label{V0}
\end{eqnarray}
The $G$ parities of the exchanged mesons $i$ are left as a $\zeta_{i}$ factor.
Since $\pi$ and $\omega$ mesons carry odd $G$ parity, $\zeta_{\pi}$, and
$\zeta_{\omega}$ should equal $-1$, and others equal $1$.

\renewcommand\tabcolsep{0.3cm}
\renewcommand{\arraystretch}{1.35}
\begin{table}[h!]
 \centering
 \caption{The flavor factors $I_e$ for charmed-strange interactions.  The values for charmed-antistrange interactions can be obtained by Eq.~(\ref{V0}) from these of charmed-strange interactions. The { $I_\sigma$} should be 0 for coupling between different channels.\label{flavor factor}}
\begin{tabular}{l|c|cccccc}\toprule[1.5pt]
&$I$& $\pi$& $\eta$ &$\rho$&$\omega$  &$\sigma$ \\\hline
$\Lambda_c\Lambda$-$\Lambda_c\Lambda$ &$0$&$-$ &$-$ &$-$ & $\sqrt2$&$2$\\
$\Sigma^{(*)}_c\Lambda$-$\Sigma^{(*)}_c\Lambda$ &$1$&$-$ &$-$ &$-$ & $\frac{1}{\sqrt2}$&$1$\\
$\Lambda_c\Sigma^{(*)}$-$\Lambda_c\Sigma^{(*)}$ &$1$&$-$ &$-$ &$-$ & $\sqrt2$&$2$\\
$\Sigma^{(*)}_c\Sigma^{(*)}$-$\Sigma^{(*)}_c\Sigma^{(*)}$& $0$&$-\sqrt2$ &$\frac{1}{\sqrt6}$&$-\sqrt2$ & $\frac{1}{\sqrt2}$&$1$\\
 &$1$&$-\frac{1}{\sqrt2}$&$\frac{1}{\sqrt6}$&$-\frac{1}{\sqrt2}$& $\frac{1}{\sqrt2}$&$1$\\
 &$2$&$\frac{1}{\sqrt2}$&$\frac{1}{\sqrt6}$&$\frac{1}{\sqrt2}$& $\frac{1}{\sqrt2}$&$1$\\
$\Lambda_c\Lambda$-$\Sigma^{(*)}_c\Sigma^{(*)}$ & $0$&$-\sqrt{3}$ &$-$ &$-\sqrt{3}$ & $-$&$-$\\
$\Sigma^{(*)}_c\Lambda$-$\Lambda_c\Sigma^{(*)}$ & $1$&$1$ &$-$ &$1$ & $-$&$-$\\
 $\Lambda_c\Sigma^{(*)}$-$\Sigma^{(*)}_c\Sigma^{(*)}$& $1$&$\sqrt{2}$ &$-$ &$\sqrt{2}$ & $-$&$-$\\
 $\Sigma^{(*)}_c\Lambda$-$\Sigma^{(*)}_c\Sigma^{(*)}$ &$1$&$-1$ &$-$ &$-1$ & $-$&$-$\\
\bottomrule[1.5pt]
\end{tabular}
\end{table}

\subsection{The qBSE approach}

The Bethe-Salpeter equation is  widely used to treat two body scattering. The
potentials obtained above can be taken as Bethe-Salpeter equation under the
ladder approximation, which  describes the interaction well.  In order to reduce
the 4-dimensional Bethe-Salpeter equation to a 3-dimensional equation, we adopt
the covariant spectator approximation, which  keeps the unitary and covariance
of the equation~\cite{Gross:1991pm}. In such treatment, one of the constituent
particles, usually heavier one, is put on shell, which leads to a reduced
propagator for two constituent particles in the center-of-mass frame
as~\cite{He:2015mja,He:2011ed},

\begin{eqnarray}
	G_0&=&\frac{\delta^+(p''^{0}_h-E_h({\rm p}''))}{2E_h({\rm p''})[(W-E_h({\rm
p}''))^2-E_l^{2}({\rm p}'')]}.
\end{eqnarray}
As required by the spectator approximation, the heavier particle  ($h$ represents the charmed baryons) satisfies $p''^0_h=E_{h}({\rm p}'')=(m_{h}^{~2}+\rm p''^2)^{1/2}$. The $p''^0_l$ for the lighter particle (remarked as $l$) is then $W-E_{h}({\rm p}'')$. Here and hereafter, the value of the momentum  in center-of-mass frame are defiend as ${\rm p}=|{\bm p}|$.

After the covariant spectator approximation, the 3-dimensional Bethe-Saltpeter
equation can be reduced to a 1-dimensional equation with fixed spin-parity $J^P$
by partial wave decomposition~\cite{He:2015mja},
\begin{eqnarray}
i{\cal M}^{J^P}_{\lambda'\lambda}({\rm p}',{\rm p})
&=&i{\cal V}^{J^P}_{\lambda',\lambda}({\rm p}',{\rm
p})+\sum_{\lambda''}\int\frac{{\rm
p}''^2d{\rm p}''}{(2\pi)^3}\nonumber\\
&\cdot&
i{\cal V}^{J^P}_{\lambda'\lambda''}({\rm p}',{\rm p}'')
G_0({\rm p}'')i{\cal M}^{J^P}_{\lambda''\lambda}({\rm p}'',{\rm
p}),\quad\quad \label{Eq: BS_PWA}
\end{eqnarray}
where the sum extends only over nonnegative helicity $\lambda''$.
The partial wave potential in 1-dimensional equation is defined with the potential
of the interaction obtained in the above as
\begin{eqnarray}
{\cal V}_{\lambda'\lambda}^{J^P}({\rm p}',{\rm p})
&=&2\pi\int d\cos\theta
~[d^{J}_{\lambda\lambda'}(\theta)
{\cal V}_{\lambda'\lambda}({\bm p}',{\bm p})\nonumber\\
&+&\eta d^{J}_{-\lambda\lambda'}(\theta)
{\cal V}_{\lambda'-\lambda}({\bm p}',{\bm p})],
\end{eqnarray}
where $\eta=PP_1P_2(-1)^{J-J_1-J_2}$ with $P$ and $J$ being parity and spin for
the system. The initial and final relative momenta are chosen as ${\bm
p}=(0,0,{\rm p})$ and ${\bm p}'=({\rm p}'\sin\theta,0,{\rm p}'\cos\theta)$. The
$d^J_{\lambda\lambda'}(\theta)$ is the Wigner d-matrix.  A regularization is
usually introduced to avoid divergence, when we treat an integral equation. In
the qBSE approach,  we usually adopt an exponential regularization by
introducing a form factor into the propagator as
$f(q^2)=e^{-(k_l^2-m_l^2)^2/\Lambda_r^4}$, where $k_l$ and $m_l$ are the
momentum and mass of the lighter  baryon.  In the current work, the
relation of the cutoff $\Lambda_r= m + \alpha_r$ 0.22~GeV with $m$ being the
mass of the exchanged meson is also introduced into the regularization form
factor as in those for the exchanged mesons. Cutoffs  $\Lambda_e$ and $\Lambda_r$
play analogous roles in the calculation of the binding energy. For
simplification, we set  $\Lambda_e=\Lambda_r$ in the calculations.

The partial-wave qBSE is a one-dimensional integral equation, which can be
solved by discretizing the momenta with the Gauss quadrature.  It leads to a
matrix equation of a form $M=V+VGM$~\cite{He:2015mja}. The molecular state
corresponds to the pole of the amplitude, which can be obtained by varying $z$
to satisfy $|1-V(z)G(z)|=0$ where $z=E_R-i\Gamma/2$ being the exact position of
the bound state.

\section{Molecular states from  charmed-strange interactions}\label{Sec:BcB}

First we consider the charmed-strange interactions with  $C$=1, $S$=$-1$.  Only
$S$-wave states are considered in single-channel calculation. The results for
scalar, vector and tensor isospins are presented in the followings.

\subsection{Isoscalar charmed-strange molecular states}

\begin{figure}[h!]
  \centering
  \includegraphics[scale=0.65,bb=407 136 14 491]{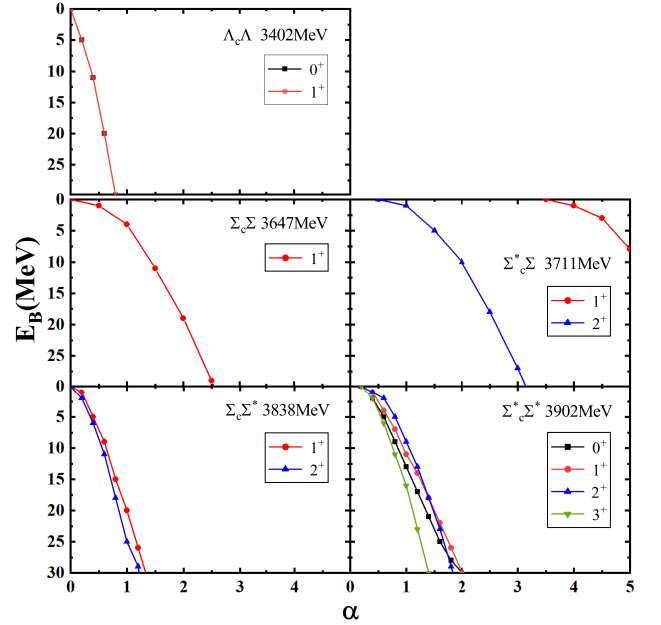}
  \caption{Binding energies of isoscalar bound states from the charmed-strange interactions  with the variation of $\alpha$ in single-channel calculation.}
  \label{0a}
\end{figure}

In the current model, we have only one free parameter $\alpha$. In
the following, we vary the free parameter in a range of 0-5 to find the bound states with
binding energy smaller than 30~MeV.  The single-channel results of
charmed-strange interaction with isospin $I=0$ are  illustrated in Fig.~\ref{0a}.

For the isoscalar charmed-strange interaction, we consider twelve channels,
$\Lambda_c\Lambda$ with spin parities $J^P=(0,1)^+$, $\Sigma_c\Sigma$ with
$(0,1)^+$, $\Sigma_c\Sigma^*$ and $\Sigma^*_c\Sigma$ with $(1,2)^+$, and
$\Sigma^*_c\Sigma^*$ with $(0,1,2,3)^+$. As shown in  Fig.~\ref{0a}, the
isoscalar single-channel calculation suggests that except $\Sigma_c\Sigma$
interaction with $0^+$, all other eleven isoscalar channels can produce bound
state. Their binding energies increase with increaseing the free parameter
$\alpha$.  The bound states from the interactions $\Lambda_c\Lambda$ appear at
$\alpha$ value of about 0, and increase rapidly to 30 MeV at  $\alpha$ value of
about 1. The bound states from  the $\Sigma_c\Sigma$ interaction with $1^+$ and
the $\Sigma^*_c\Sigma$ interaction with $2^+$ also appear at an $\alpha$ of
about 1, and increase relatively slowly to 30~MeV at $\alpha$ value of about 3.
The attraction of the $\Sigma^*_c\Sigma$ interaction with $1^+$ is weak, and
produce a bound state at an $\alpha$ value of about $3.5$.  The variation
tendencies of the binding energies of two states with different spin parities
from the $\Lambda_c\Lambda$ interaction are analogous.

The channels with the same quantum numbers can couple to each other, which will
make the poles move in the complex energy plane. In Table~\ref{t0a}, we present
the coupled-channel results of isoscalar charmed-strange interactions. The
results of pole under the corresponding threshold with different $\alpha$ is
given in the second and third columns with full coupled-channel interaction. To
compare with the single-channel results, we present the position as $M_{th}-z$
instead of the position  $z$ of the pole, with the $M_{th}$ being the nearest
threshold.  The results are similar to those from the single-channel
calculations.  For the states above the lowest threshold, with the
coupled-channel effect, the pole of a bound state will deviate from the real
axis and acquire an imaginary part, which corresponds to the width as $\Gamma=-2
{\rm Im} z$. The results suggest small width produced from the couplings with
the channels considered.

\renewcommand\tabcolsep{0.141cm}
\renewcommand{\arraystretch}{1.35}
\begin{table}[h!]
\footnotesize
\caption{The masses and widths of isoscalar charmed-strange molecular states at different values of $\alpha$. The ``$CC$" means full coupled-channel calculation.  The values of the complex position means mass of corresponding threshold subtracted by the position of a pole, $M_{th}-z$,  in the unit of MeV. The two short line "$--$" means the decay channel does not exist. The imaginary part shown as "$0.0$" means too small value under the current precision chosen.\label{t0a}}
\begin{tabular}{crrrrrrrrrr}\bottomrule[1.5pt]
\specialrule{0em}{1pt}{1pt}
$I=0$&$\alpha$ &\multicolumn{1}{c}{$CC$}& \multicolumn{1}{c}{$\Sigma_c\Sigma^*$} & \multicolumn{1}{c}{$\Sigma^*_c\Sigma$} &   \multicolumn{1}{c}{$\Sigma_c\Sigma$} &  \multicolumn{1}{c}{  $\Lambda_c\Lambda$} \\
\specialrule{0em}{1pt}{1pt}
\hline
\specialrule{0em}{1pt}{1pt}
$\Sigma^*_c\Sigma^*(0^+)$
&$0.5$ &$3+0.2i$   &$3+0.0i$     &$3+0.0i$      &$3+0.1i$  &$3+0.1i$      \\
3902~MeV
&$1.0$ &$9+0.5i$   &$13+0.0i$    &$13+0.1i$      &$13+0.6i$  &$9+0.6i$     \\
&$1.5$ &$10+8.9i$   &$24+0.0i$    &$23+0.1i$       &$24+1.8i$  &$12+1.0i$     \\
\specialrule{0em}{1pt}{1pt}
$\Sigma^*_c\Sigma^*(1^+)$
&$0.5$ &$2+0.3i$   &$2+0.1i$     &$2+0.1i$      &$2+0.1i$  &$2+0.0i$      \\
3902~MeV
&$1.0$ &$8+2.0i$   &$10+0.2i$    &$10+0.4i$      &$10+0.5i$  &$8+0.3i$     \\
&$1.5$ &$13+7.5i$   &$20+0.4i$    &$20+1.5i$       &$20+1.2i$  &$15+0.8i$     \\
\specialrule{0em}{1pt}{1pt}
$\Sigma^*_c\Sigma^*(2^+)$
&$0.5$ &$1+0.2i$   &$1+0.1i$     &$1+0.0i$      &$1+0.0i$  &$1+0.0i$      \\
3902~MeV
&$1.0$ &$7+1.6i$   &$6+1.3i$    &$7+0.4i$      &$8+0.2i$  &$8+0.2i$     \\
&$1.5$ &$12+3.6i$   &$14+1.5i$    &$20+1.8i$       &$20+1.0i$  &$17+0.6i$     \\
\specialrule{0em}{1pt}{1pt}
$\Sigma^*_c\Sigma^*(3^+)$
&$0.5$ &$3+0.3i$   &$3+0.2i$     &$3+0.0i$      &$3+0.0i$  &$3+0.0i$      \\
3902~MeV
&$1.0$ &$15+1.8i$   &$17+2.0i$    &$16+0.3i$      &$16+0.2i$  &$14+0.3i$     \\
&$1.5$ &$27+4.0i$   &$30+4.5i$    &$39+3.0i$       &$32+1.6i$  &$27+1.2i$     \\
\specialrule{0em}{1pt}{1pt}
$\Sigma_c\Sigma^*(1^+)$
&$0.3$ &$3+0.1i$   &$--$     &$3+0.0i$      &$3+0.1i$  &$3+0.0i$      \\
3838~MeV
&$0.5$ &$6+0.3i$   &$--$    &$7+0.1i$      &$7+0.2i$  &$6+0.1i$     \\
&$0.7$ &$11+0.7i$   &$--$    &$12+0.2i$       &$12+0.3i$  &$11+0.2i$     \\
\specialrule{0em}{1pt}{1pt}
$\Sigma_c\Sigma^*(2^+)$
&$0.3$ &$3+0.1i$   &$--$  &$3+0.0i$      &$1+0.0i$  &$3+0.0i$      \\
3838~MeV
&$0.5$ &$8+0.2i$&$--$    &$8+0.0i$      &$3+0.0i$  &$8+0.0i$     \\
&$0.7$ &$14+0.4i$&$--$    &$14+0.1i$       &$8+0.1i$  &$14+0.1i$     \\
\specialrule{0em}{1pt}{1pt}
$\Sigma^*_c\Sigma(1^+)$
&$3.5$ &$0+2.1i$   &$--$     &$--$      &$0+0.2i$  &$1+5.6i$      \\
3711~MeV
&$3.7$ &$1+2.4i$   &$--$    &$--$      &$1+0.3i$  &$3+4.4i$     \\
&$4.3$ &$3+3.3i$   &$--$    &$--$       &$2+0.3i$  &$14+9.6i$     \\
\specialrule{0em}{1pt}{1pt}
$\Sigma^*_c\Sigma(2^+)$
&$1.0$ &$1+0.1i$   &$--$     &$--$      &$1+0.0i$  &$1+0.0i$      \\
3711~MeV
&$1.5$ &$4+0.4i$   &$--$    &$--$      &$5+0.2i$  &$4+0.4i$     \\
&$2.0$ &$8+0.8i$   &$--$    &$--$       &$10+0.3i$  &$8+0.5i$     \\
\specialrule{0em}{1pt}{1pt}
$\Sigma_c\Sigma(1^+)$
&$0.5$ &$1+0.0i$   &$--$     &$--$      &$--$  &$1+0.0i$      \\
3647~MeV
&$1.0$ &$4+0.0i$   &$--$    &$--$      &$--$  &$4+0.0i$     \\
&$1.5$ &$10+0.2i$   &$--$    &$--$       &$--$  &$10+0.2i$     \\
\bottomrule[1.5pt]
\end{tabular}
\end{table}

The two-channel calculations are also performed to show the strength of the
coupling between the molecular states and the corresponding decay channels. Larger
variation of the mass and value of width reflect stronger couplings.  In the
fourth to seventh columns, the results for the couplings to labeled channels are
presented. For the state with $(0,1)^+$ near $\Sigma_c^*\Sigma{^*}$ thresholds, the
strong couplings to the $\Sigma_c\Sigma$ channel can be found based on the mass
and width, while the states with $(2,3)^+$ couples strongly to the
$\Sigma_c\Sigma^*$ channel. No obvious strongly coupled channel can be found for
the states near the $\Sigma_c\Sigma^*$,  $\Sigma_c^*\Sigma$ and  $\Sigma_c\Sigma$
thresholds. Since the $\Lambda_c\Lambda$ interaction has the lowest threshold,
no width will be acquired from the coupled-channel calculation, and the results are not presented due to absence of the decay channels.

\subsection{Isovector charmed-strange molecular states}

For  isovector charmed-strange interaction states, we consider eighteen
channels, $\Lambda_c\Sigma$ , $\Sigma_c\Lambda$ and
$\Sigma_c\Sigma$ with $(0,1)^+$, $\Sigma^*_c\Lambda$, $\Lambda_c\Sigma^*$,
$\Sigma^*_c\Sigma$ and $\Sigma_c\Sigma^*$ with $(1,2)^+$, and
$\Sigma^*_c\Sigma^*$ with $(0,1,2,3)^+$. The results are shown in the following
Fig.~\ref{1a}.

\begin{figure}[h!]
  \centering
  \includegraphics[scale=0.65,bb=401 25 23 497,clip]{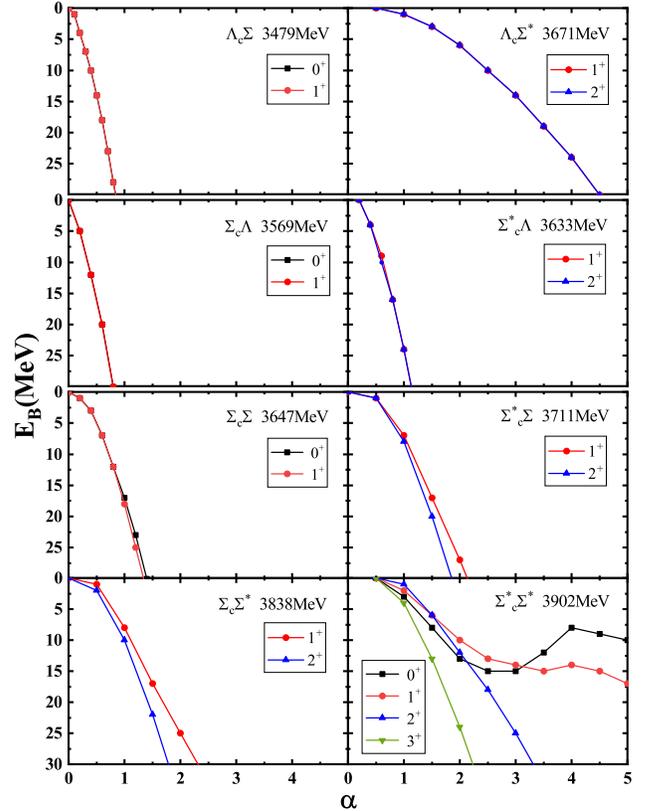}
  \caption{Binding energies of isovector bound states from the charmed-strange interactions  with the variation of $\alpha$ in single-channel calculation.}
  \label{1a}
\end{figure}

From Fig.~\ref{1a}, the results suggest that bound states are produced from all
eighteen channels and all  bound states are produced at a value of $\alpha$ less
than $1$.  It is still worth mentioning that at a $\alpha$ value of about $4$ or
larger, the repulsion from the $\omega$ meson and $\eta$ meson exchanges
increase faster than  attractions of other mesons, which make the states with
$0^+$ and $1^+$ of $\Sigma^*_c\Sigma^*$ shallower. Also, the variation
tendencies of the binding energies of states with different spin parities from
$\Lambda_c\Sigma^{(*)}$ or $\Sigma^{(*)}\Lambda$ interaction are still analogous. One can generally find that  the binding energies of states
with the smaller spin increase more rapidly for the isovector bound states.

\renewcommand\tabcolsep{0.39cm}
\renewcommand{\arraystretch}{1.31}
\begin{table*}[hpbt!]   
\footnotesize
\caption{The masses and widths of isovector charmed-strange molecular states at different values of $\alpha$. Other notations are the same as Table~\ref{t0a}.\label{t1a}}
\begin{tabular}{crrrrrrrrrrrrrrrr}\bottomrule[1.5pt]
\specialrule{0em}{1pt}{1pt}
$I=1$&$\alpha$ &\multicolumn{1}{c}{$CC$}& \multicolumn{1}{c}{$\Sigma_c\Sigma^*$} & \multicolumn{1}{c}{$\Sigma^*_c\Sigma$} & \multicolumn{1}{c}{$\Lambda_c\Sigma^*$} &   \multicolumn{1}{c}{$\Sigma_c\Sigma$} &\multicolumn{1}{c}{$\Sigma^*_c\Lambda$}&\multicolumn{1}{c}{$\Sigma_c\Lambda$} &\multicolumn{1}{c}{  $\Lambda_c\Sigma$} \\
\specialrule{0em}{1pt}{1pt}
\hline
\specialrule{0em}{1pt}{1pt}
$\Sigma^*_c\Sigma^*(0^+)$
&$1.0$ &$3+3.7i$   &$3+0.0i$     &$3+0.0i$      &$1+4.1i$  &$3+0.0i$  &$3+0.3i$ &$2+0.7i$&$3+0.1i$   \\
3902 {\rm MeV}
&$1.5$ &$7+5.4i$   &$8+0.3i$    &$8+0.0i$            &$5+1.0i$  &$8+0.0i$    &$7+1.1i$  &$5+3.2i$&$6+0.6i$    \\
&$2.0$ &$13+13.4i$   &$12+1.0i$    &$13+0.0i$         &$10+2.7i$  &$13+0.0i$  &$10+2.4i$ &$8+10.2i$&$8+1.4i$   \\
\specialrule{0em}{1pt}{1pt}
$\Sigma^*_c\Sigma^*(1^+)$
&$1.0$&$0+1.1i$ &$2+0.0i$   &$2+0.0i$     &$1+1.2i$      &$2+0.0i$    &$1+0.6i$      &$1+0.4i$  &$1+0.1i$    \\
3902 {\rm MeV}
&$1.5$ &$5+5.7i$&$6+0.3i$   &$6+0.0i$    &$5+5.4i$      &$6+0.0i$    &$3+2.5i$      &$3+1.7i$  &$4+0.4i$   \\
&$2.0$ &$9+13.1i$&$10+10.9i$   &$10+0.0i$    &$9+18.7i$       &$10+0.0i$     &$6+6.6i$      &$5+5.3i$  &$7+0.9i$   \\
\specialrule{0em}{1pt}{1pt}
$\Sigma^*_c\Sigma^*(2^+)$
&$1.0$ &$1+1.2i$   &$1+0.0i$      &$1+0.0i$  &$1+0.4i$ &$1+0.0i$         &$1+0.5i$   &$1+0.2i$  &$1+0.0i$    \\
3902 {\rm MeV}
&$1.5$ &$5+3.3i$   &$6+0.2i$      &$6+0.0i$  &$5+2.1i$ &$6+0.0i$        &$4+2.1i$  &$4+0.9i$  &$5+0.2i$     \\
&$2.0$ &$6+7.1i$  &$11+0.8i$      &$12+0.1i$  &$12+8.2i$  &$12+0.0i$        &$8+5.4i$ &$7+2.6i$  &$10+0.5i$      \\
\specialrule{0em}{1pt}{1pt}
$\Sigma^*_c\Sigma^*(3^+)$
&$1.0$ &$4+2.0i$   &$4+0.1i$     &$3+0.0i$      &$3+0.9i$  &$3+0.0i$ &$4+0.7i$      &$4+0.3i$  &$4+0.1i$       \\
3902 {\rm MeV}
&$1.5$ &$5+3.4i$   &$13+0.4i$    &$13+0.0i$      &$11+2.2i$  &$13+0.0i$  &$10+2.8i$      &$11+1.6i$  &$12+0.4i$     \\
&$2.0$ &$13+9.2i$   &$24+0.6i$    &$24+0.1i$       &$23+5.1i$  &$24+0.0i$ &$11+6.8i$      &$17+4.6i$  &$19+1.4i$   \\
\specialrule{0em}{1pt}{1pt}
$\Sigma_c\Sigma^*(1^+)$
&$0.5$ &$1+0.5i$   &$--$     &$1+0.0i$      &$1+0.3$  &$1+0.0i$&$1+0.2i$      &$1+0.1i$  &$1+0.0i$       \\
3838 {\rm MeV}
&$1.0$ &$7+3.1i$   &$--$    &$8+0.0i$      &$8+2.2i$  &$8+0.0i$ &$7+1.6i$      &$7+1.1i$  &$7+0.2i$     \\
&$1.5$ &$16+8.6i$   &$--$    &$16+0.0i$       &$17+9.4i$  &$17+0.0i$  &$14+5.8i$      &$14+4.6i$  &$15+0.9i$    \\
\specialrule{0em}{1pt}{1pt}
$\Sigma_c\Sigma^*(2^+)$
&$0.5$ &$1+0.3i$&$--$&$2+0.0i$&$2+0.1i$   &$2+0.0i$     &$2+0.1i$      &$2+0.1i$  &$2+0.0i$      \\
3838 {\rm MeV}
&$1.0$ &$8+1.7i$&$--$&$10+0.0i$&$10+0.9i$   &$10+0.0i$    &$10+0.8i$      &$9+0.7i$  &$10+0.1i$     \\
&$1.5$ &$17+4.9i$&$--$ &$22+0.0i$&$20+2.3i$  &$22+0.0i$    &$21+2.7i$       &$17+2.1i$  &$21+0.6i$     \\
\specialrule{0em}{1pt}{1pt}
$\Sigma^*_c\Sigma(1^+)$
&$0.5$ &$0+0.4i$   &$--$     &$--$      &$0+0.2i$  &$1+0.0i$ &$1+0.1i$  &$1+0.1i$     &$1+0.0i$      \\
3711 {\rm MeV}
&$1.0$ &$7+1.1i$   &$--$    &$--$      &$7+0.5i$  &$7+0.0i$ &$7+0.5i$  &$7+0.4i$     &$7+0.2i$     \\
&$1.5$ &$18+2.2i$   &$--$    &$--$       &$17+0.5i$  &$17+0.0i$ &$18+1.3i$   &$16+1.5i$     &$17+0.8i$    \\
\specialrule{0em}{1pt}{1pt}
$\Sigma_c\Sigma^*(2^+)$
&$0.5$ &$1+0.4i$&$--$&$--$  &$1+0.4i$&$1+0.0i$&$1+0.0i$ &$1+0.0i$&$1+0.0i$    \\
3711 {\rm MeV}
&$1.0$ &$8+1.4i$&$--$&$--$   &$8+1.6i$&$8+0.0i$&$8+0.2i$ &$8+0.3i$&$8+0.2i$   \\
&$1.5$ &$17+3.5i$ &$--$&$--$   &$18+2.9i$&$20+0.1i$&$19+0.4i$ &$20+0.6i$&$18+0.7i$  \\
\specialrule{0em}{1pt}{1pt}
$\Lambda_c\Sigma^*(1^+)$
&$1.4$ &$1+1.8i$   &$--$     &$--$      &$--$  &$2+0.5i$ &$1+2.5i$  &$1+0.4i$     &$3+0.1i$      \\
3671 {\rm MeV}
&$1.7$ &$2+2.2i$   &$--$    &$--$      &$--$  &$3+0.9i$ &$3+3.0i$  &$2+1.2i$     &$6+0.2i$     \\
&$2.1$ &$3+2.6i$   &$--$    &$--$       &$--$  &$6+1.7i$ &$6+3.9i$   &$3+2.0i$     &$14+1.0i$    \\
\specialrule{0em}{1pt}{1pt}
$\Lambda_c\Sigma^*(2^+)$
&$1.5$ &$4+1.9i$   &$--$     &$--$      &$--$  &$3+0.1i$ &$2+1.7i$  &$3+0.1i$     &$3+0.1i$      \\
3671 {\rm MeV}
&$2.0$ &$6+3.8i$   &$--$    &$--$      &$--$  &$7+0.1i$ &$5+2.9i$  &$8+0.6i$     &$6+0.2i$     \\
&$2.5$ &$8+5.4i$   &$--$    &$--$       &$--$  &$11+1.1i$ &$7+4.7i$   &$12+1.9i$     &$10+0.5i$    \\
\specialrule{0em}{1pt}{1pt}
$\Sigma_c\Sigma(0^+)$
&$0.2$ &$1+0.0i$   &$--$     &$--$      &$--$  &$--$  &$1+0.0i$ &$1+0.0i$  &$1+0.0i$      \\
3647 {\rm MeV}
&$0.4$ &$3+0.1i$   &$--$    &$--$      &$--$  &$--$ &$3+0.0i$ &$3+0.1i$  &$3+0.0i$      \\
&$0.6$ &$7+0.1i$   &$--$    &$--$       &$--$  &$--$   &$7+0.0i$ &$7+0.1i$  &$7+0.0i$    \\
\specialrule{0em}{1pt}{1pt}
$\Sigma_c\Sigma(1^+)$
&$0.2$ &$0+0.0i$   &$--$     &$--$      &$--$  &$--$  &$0+0.0i$ &$1+0.0i$  &$1+0.0i$      \\
3647 {\rm MeV}
&$0.4$ &$4+0.0i$   &$--$    &$--$      &$--$  &$--$ &$4+0.0i$ &$3+0.1i$  &$3+0.0i$      \\
&$0.6$ &$10+0.0i$   &$--$    &$--$       &$--$  &$--$   &$10+0.0i$ &$7+0.2i$  &$7+0.0i$    \\
\specialrule{0em}{1pt}{1pt}
$\Sigma^*_c\Lambda(1^+)$
&$0.3$ &$2+0.0i$   &$--$     &$--$      &$--$  &$--$  &$--$ &$2+0.0i$  &$2+0.0i$      \\
3633 {\rm MeV}
&$0.5$ &$7+0.0i$   &$--$    &$--$      &$--$  &$--$ &$--$ &$7+0.0i$  &$7+0.2i$      \\
&$0.7$ &$13+0.0i$   &$--$    &$--$       &$--$  &$--$   &$--$ &$13+0.0i$  &$13+0.4i$    \\
\specialrule{0em}{1pt}{1pt}
$\Sigma^*_c\Lambda(2^+)$
&$0.3$ &$2+0.0i$   &$--$     &$--$      &$--$  &$--$  &$--$ &$2+0.0i$  &$2+0.0i$      \\
3633 {\rm MeV}
&$0.5$ &$7+0.1i$   &$--$    &$--$      &$--$  &$--$ &$--$ &$7+0.0i$  &$7+0.1i$      \\
&$0.7$ &$13+0.2i$   &$--$    &$--$       &$--$  &$--$   &$--$ &$13+0.0i$  &$7+0.2i$    \\
\specialrule{0em}{1pt}{1pt}
$\Sigma_c\Lambda(0^+)$
&$0.2$ &$3+0.0i$   &$--$     &$--$      &$--$  &$--$  &$--$ &$--$  &$3+0.0i$      \\
3569 {\rm MeV}
&$0.3$ &$6+0.0i$   &$--$    &$--$      &$--$  &$--$ &$--$ &$--$  &$6+0.0i$      \\
&$0.5$ &$14+0.0i$   &$--$    &$--$       &$--$  &$--$   &$--$ &$--$  &$14+0.0i$    \\
\specialrule{0em}{1pt}{1pt}
$\Sigma_c\Lambda(1^+)$
&$0.1$ &$2+0.0i$   &$--$     &$--$      &$--$  &$--$  &$--$ &$--$  &$2+0.0i$      \\
3569 {\rm MeV}
&$0.3$ &$8+0.0i$   &$--$    &$--$      &$--$  &$--$ &$--$ &$--$  &$8+0.0i$      \\
&$0.5$ &$16+0.1i$   &$--$    &$--$       &$--$  &$--$   &$--$ &$--$  &$16+0.1i$    \\
\specialrule{0em}{1pt}{1pt}
\bottomrule[1.5pt]
\end{tabular}
\end{table*}

In Table~\ref{t1a}, we  present the coupled-channel results of isovector
charmed-strange interactions.  There are four isovector states near the
$\Sigma^*_c\Sigma^*$ threshold.  The width for the $0^+$ state is mainly from
the $\Sigma_c\Lambda$ channel.  For the state with $1^+$, large couplings can be
found in the channels $\Lambda_c\Sigma^*$ and $\Sigma_c\Sigma^*$.  The state
with $2^+$ strongly couples to the channels $\Lambda_c\Sigma^*$ and
$\Sigma_c^*\Lambda$. The states with $3^+$ has strong couplings to the channels
$\Sigma^*_c\Lambda$, $\Lambda_c\Sigma^*$, and $\Sigma_c\Lambda$.  Near the
$\Sigma_c\Sigma^*$ threshold, there exist two poles with spin parities $1^+$ and
$2^+$. For the state with $1^+$, the $\Lambda_c\Sigma^*$ is found to be its
dominant decay channel while no obvious dominant channel can be found for the
state with $2^+$.  For two states near the $\Sigma^*_c\Sigma$ threshold, there
is also no obvious dominant channel for the $1^+$ states while the $2^+$ state
has the strongest coupling to channel $\Lambda_c\Sigma^*$.  For two
$\Lambda_c\Sigma^*$ states, the dominant channel shifts to the
$\Sigma^*_c\Lambda$ channel. For other states, the widths are very small partly
due to few channels below their masses.

\subsection{Isotensor molecular states}

\begin{figure}[h!]
  \centering
  \includegraphics[scale=0.65,bb=398 248 33 494]{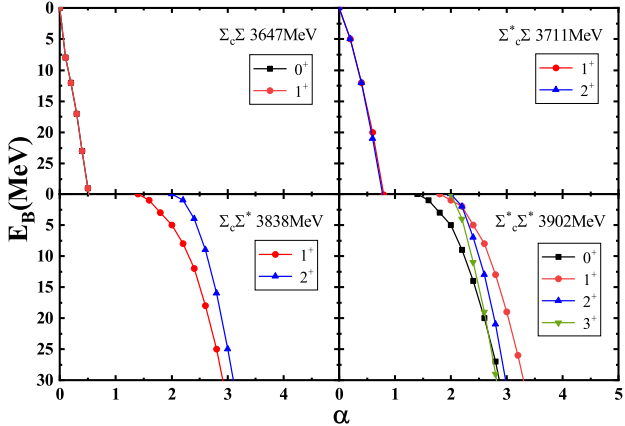}
  \caption{Binding energies of isotensor bound states from the charmed-strange interactions  with the variation of $\alpha$ in single-channel calculation.}
  \label{2a}
\end{figure}

In Fig.~\ref{2a}, the isotensor bound states from the charmed-strange
interactions in a single-channel calculation are presented. Channels
considered include $\Sigma_c\Sigma$ with $(0,1)^+$,  $\Sigma^*_c\Sigma$ and
$\Sigma_c\Sigma^*$ with $(1,2)^+$, and $\Sigma^*_c\Sigma^*$ with $(0,1,2,3)^+$.
In the considered range of parameter $\alpha$, bound states are produced from
all channels. However, two  $\Sigma_c\Sigma$ and $\Sigma^*_c\Sigma$ bound states
with $(0,1)^+$ appears at $\alpha$ values of about 0, and increase rapidly to 30
MeV at $\alpha$ values less than 1 while the states from the interactions
$\Sigma_c\Sigma^*$ and $\Sigma^*_c\Sigma^*$ appears at an $\alpha$ values of 1.5
or larger and reach 30~MeV at $\alpha$ values about 3. Such results disfavor
the coexistence of the these states.

In Table.~\ref{t2a}, we  present the coupled-channel results of isotensor
charmed-strange interactions. Though the overall coupled-channel results with
all four channels, the mass of the $\Sigma^*_c\Sigma^*$ state with the $3^+$
state becomes obviously smaller, which leads to a larger parameter $\alpha$ to
produce the state, about $2.5$, compared with the single-channel calculation in
Fig.~\ref{2a}, which is mainly from coupling to the $\Sigma^*_c\Sigma$ channel.
From the two-channel calculations listed in the fourth to sixth columns, the
isotensor $\Sigma^*_c\Sigma^*$ states with $(0,1)^+$ have the strongest coupling
to the $\Sigma_c\Sigma$ channel, while states with $(2,3)^+$ states prefer the
$\Sigma_c\Sigma^*$ channel. For the $\Sigma_c\Sigma^*$ states with $(1,2)^+$, the
$\Sigma^*_c\Sigma$ channel is dominant to produce their total width. The
coupling effect has no effect on the width for the $\Sigma^*_c\Sigma$ states with
$(1,2)^+$, while the degeneration in mass disappears for the two states.

\renewcommand\tabcolsep{0.22cm}
\renewcommand{\arraystretch}{1.45}
\begin{table}[hpbt!]   
\footnotesize
\caption{The masses and widths of isotensor charmed-strange  molecular states at different values of $\alpha$. Other notations are the same as Table~\ref{t0a}.\label{t2a}}
\begin{tabular}{crrrrrrrr}\bottomrule[1.5pt]
\specialrule{0em}{1pt}{1pt}
$I=2$&$\alpha$ &\multicolumn{1}{c}{$CC$}& \multicolumn{1}{c}{$\Sigma_c\Sigma^*$} & \multicolumn{1}{c}{$\Sigma^*_c\Sigma$} &   \multicolumn{1}{c}{$\Sigma_c\Sigma$}  \\
\specialrule{0em}{1pt}{1pt}
\hline
\specialrule{0em}{1pt}{1pt}
$\Sigma^*_c\Sigma^*(0^+)$
&$1.5$ &$0+1.0i$   &$1+0.0i$     &$0+0.1i$      &$0+0.5i$       \\
3902 {\rm MeV}
&$2.0$ &$5+4.6i$   &$8+0.1i$    &$4+0.6i$      &$3+3.3i$      \\
&$2.5$ &$19+9.3i$   &$22+0.1i$    &$14+1.5i$       &$13+10.2i$      \\
\specialrule{0em}{1pt}{1pt}
$\Sigma^*_c\Sigma^*(1^+)$
&$1.7$ &$1+0.7i$   &$2+0.4i$     &$0+0.2i$      &$0+0.2i$        \\
3902 {\rm MeV}
&$2.0$ &$5+1.5i$   &$6+1.1i$    &$2+1.6i$      &$1+1.6i$      \\
&$2.5$ &$18+2.8i$   &$22+3.6i$    &$9+4.7i$       &$9+5.6i$     \\
\specialrule{0em}{1pt}{1pt}
$\Sigma^*_c\Sigma^*(2^+)$
&$2.2$ &$0+1.9i$       &$1+0.7i$  &$2+1.2i$&$2+0.0i$        \\
3902 {\rm MeV}
&$2.5$ &$3+3.2i$       &$6+1.7i$  &$8+1.9i$  &$10+0.0i$     \\
&$3.0$ &$21+5.8i$         &$16+5.7i$  &$25+3.6i$&$29+0.0i$       \\
\specialrule{0em}{1pt}{1pt}
$\Sigma^*_c\Sigma^*(3^+)$
&$2.2$ &$0+6.6i$   &$4+5.3i$     &$0+2.4i$      &$4+0.0i$        \\
3902 {\rm MeV}
&$2.5$ &$1+14.2i$   &$14+9.0i$    &$8+5.2i$      &$14+0.0i$       \\
&$2.8$ &$7+18.0i$   &$27+11.8i$    &$15+8.4i$       &$28+0.2i$  \\
\specialrule{0em}{1pt}{1pt}
$\Sigma_c\Sigma^*(1^+)$
&$1.5$ &$0+0.9i$   &$--$          &$1+0.8i$  &$0+0.6i$      \\
3838 {\rm MeV}
&$2.0$ &$5+4.1i$   &$--$        &$6+3.6i$  &$5+1.9i$     \\
&$2.5$ &$15+8.0i$   &$--$         &$21+8.5i$  &$15+3.3i$     \\
\specialrule{0em}{1pt}{1pt}
$\Sigma_c\Sigma^*(2^+)$
&$2.2$ &$2+5.6i$   &$--$     &$2+5.2i$      &$0+0.2i$        \\
3838 {\rm MeV}
&$2.4$ &$9+8.4i$   &$--$    &$9+7.7i$      &$2+1.0i$      \\
&$2.6$ &$15+9.4i$   &$--$    &$18+9.2i$       &$15+3.1i$       \\
\specialrule{0em}{1pt}{1pt}
$\Sigma^*_c\Sigma(1^+)$
&$0.1$ &$2+0.0i$   &$--$     &$--$      &$2+0.0i$        \\
3711 {\rm MeV}
&$0.3$ &$8+0.1i$   &$--$    &$--$      &$8+0.1i$       \\
&$0.5$ &$16+0.3i$   &$--$    &$--$       &$16+0.3i$       \\
\specialrule{0em}{1pt}{1pt}
$\Sigma^*_c\Sigma(2^+)$
&$0.1$ &$2+0.0i$   &$--$     &$--$      &$2+0.0i$      \\
3711 {\rm MeV}
&$0.3$ &$8+0.1i$   &$--$    &$--$      &$8+0.1i$     \\
&$0.5$ &$10+0.4i$   &$--$    &$--$       &$10+0.4i$     \\
\bottomrule[1.5pt]
\end{tabular}
\end{table}

\section{Molecular states from  charmed-antistrange interactions}\label{Sec:BcBbar}

Now we turn to the charmed-antistrange systems with $C$=1 and $S$=$1$ by replacing
the strange baryons with their antibaryons in the corresponding systems
discussed above by the $G$ parity rule. We still  consider $S$-wave states with
scalar, vector and tensor isospins in single-channel calculation. The
coupled-channel calculation with all channels and two channels will be
performed, and the isoscalar, isovector and isotensor results will be shown  in
the followings.

\subsection{Isoscalar molecular states}

\begin{figure}[h!]
   \centering
   \includegraphics[scale=0.65,bb=400 352 23 500,clip]{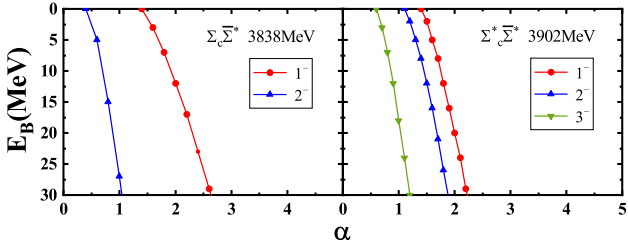}
   \caption{Binding energies of isoscalar bound states from the charmed-antistrange interactions  with the variation of $\alpha$ in single-channel calculation.}
   \label{0b}
 \end{figure}

For the isoscalar charmed-antistrange interaction, there also exist twelve
channels, $\Lambda_c\bar{\Lambda}$ with spin parities $J^P=(0,1)^-$,
$\Sigma_c\bar{\Sigma}$ with $(0,1)^-$, $\Sigma_c\bar{\Sigma}^*$ and
$\Sigma^*_c\bar{\Sigma}$ with $(1,2)^-$, and $\Sigma^*_c\bar{\Sigma}^*$ with
$(0,1,2,3)^-$.   Compared with the charmed-strange interactions, where eleven
bound states are produced, only five states were produced here from the twelve
channels as shown in Fig.~\ref{0b}.

The $\Sigma_c\bar{\Sigma}^*$ and $\Sigma^*_c\bar{\Sigma}^*$ interactions are
found attractive and produce two and three bound states with spin parities
$(1,2)^-$ and $(1,2,3)^-$, respectively. The $\Sigma_c\bar{\Sigma}^*$ states
with $(1,2)^-$ appear at  $\alpha$ values about 0.5 and 1.5 and the
$\Sigma^*_c\bar{\Sigma}^*$ states with $(1,2,3)^-$ appear at $\alpha$ values of
about 0.5, 1.0 and 1.5, respectively.  At the same time, the larger isospin
quantum number, easier it is to attract for the states. The phenomenon is the
same as what we draw in the above isovector charmed-strange sector. There is no
bound states for the channels $\Lambda_c\bar{\Lambda}$, $\Sigma_c\bar{\Sigma}$  and
$\Sigma^*_c\bar{\Sigma}$. It is mainly due to the different signs of flavor
factors for the $\pi$ and $\omega$ meson exchanges in Table.~\ref{flavor factor}
according to the  $G$-parity rule ~\cite{Phillips:1967zza,Klempt:2002ap}, which
means that the attraction and repulsion are opposite for these two exchanges.

The coupled-channel results are presented in Table~\ref{t0b}. The mass of bound
states $\Sigma^*_c\bar{\Sigma}^*$ with $(1,2,3)^-$ decrease obviously after
including all channels. And these states acquire widths of several to tens MeV
with binding energies of several MeV, which is mainly from the
$\Sigma_c\bar{\Sigma}^*$ channel. For the states with $(2,3)^-$, considerable
large couplings can be found in the channel $\Sigma_c^*\bar{\Sigma}$.   For two
states with $(1,2)^-$ near the $\Sigma_c\bar{\Sigma}^*$ threshold, the strongest
coupling channel is  the $\Sigma_c\bar{\Sigma}$ channel while large couplings
are also found in the channels  $\Sigma_c^*\bar{\Sigma}$ and
$\Lambda_c\bar{\Lambda}$ for state with $1^-$.

\renewcommand\tabcolsep{0.12cm}
\renewcommand{\arraystretch}{1.36}
\begin{table}[hpbt!]   
\footnotesize
\caption{The masses and widths of  isoscalar charmed-antistrange molecular states at different values of $\alpha$. Other notations are the same as Table~\ref{t0a}.\label{t0b}}
\begin{tabular}{crrrrrrrrrr}\bottomrule[1.5pt]
\specialrule{0em}{1pt}{1pt}
$I=0$&$\alpha$ &\multicolumn{1}{c}{$CC$}& \multicolumn{1}{c}{$\Sigma_c\bar{\Sigma}^*$} & \multicolumn{1}{c}{$\Sigma^*_c\bar{\Sigma}$} &   \multicolumn{1}{c}{$\Sigma_c\bar{\Sigma}$} &  \multicolumn{1}{c}{  $\Lambda_c\bar{\Lambda}$} \\
\specialrule{0em}{1pt}{1pt}
\hline
\specialrule{0em}{1pt}{1pt}
$\Sigma^*_c\bar{\Sigma}^*(1^-)$
&$1.5$ &$0+6.0i$   &$0+8.2i$     &$2+0.2i$      &$2+0.1i$  &$0+0.2i$      \\
3902~MeV
&$1.6$ &$3+6.6i$   &$5+9.4i$    &$5+0.2i$      &$5+0.1i$  &$2+0.4i$     \\
&$1.7$ &$6+6.9i$   &$10+10.2i$    &$9+0.2i$       &$8+0.1i$  &$5+0.7i$     \\
\specialrule{0em}{1pt}{1pt}
$\Sigma^*_c\bar{\Sigma}^*(2^-)$
&$1.5$ &$0+12.0i$  &$0+5.1i$ &$11+3.3i$     &$12+0.1i$      &$10+0.5i$      \\
3902~MeV
&$1.6$ &$3+12.9i$  &$2+6.2i$ &$15+3.9i$    &$16+0.1i$      &$13+0.8i$     \\
&$1.7$ &$5+13.6i$   &$5+7.2i$  &$19+4.6i$    &$21+0.1i$     &$16+1.2i$     \\
\specialrule{0em}{1pt}{1pt}
$\Sigma^*_c\bar{\Sigma}^*(3^-)$
&$0.7$ &$2+3.1$   &$3+1.9i$     &$3+1.4i$      &$3+0.0i$  &$3+0.1i$      \\
3902~MeV
&$0.8$ &$6+4.5i$   &$7+2.8i$    &$7+2.1i$      &$7+0.0i$  &$7+0.2i$     \\
&$0.9$ &$10+6.0i$   &$12+3.9 i$    &$12+2.9i$       &$12+0.0i$  &$12+0.4i$     \\
\specialrule{0em}{1pt}{1pt}
$\Sigma_c\bar{\Sigma}^*(1^-)$
&$1.5$ &$0+11.0i$   &$--$     &$2+2.9i$    &$4+4.9i$  &$0+2.9i$      \\
3838~MeV
&$1.6$ &$2+13.0i$   &$--$    &$3+3.9i$     &$6+5.8i$  &$2+3.6i$     \\
&$1.7$ &$4+15.0i$   &$--$    &$5+4.7i$     &$9+6.6i$  &$4+4.4i$     \\
\specialrule{0em}{1pt}{1pt}
$\Sigma_c\bar{\Sigma}^*(2^-)$
&$0.5$ &$1+1.3i$   &$--$     &$1+0.5i$      &$1+0.9i$  &$1+0.0i$      \\
3838~MeV
&$0.6$ &$5+2.1i$   &$--$    &$5+0.7i$      &$5+1.4i$  &$5+0.0i$     \\
&$0.7$ &$10 +3.0i$   &$--$    &$10+1.0 i$       &$9+2.0i$  &$10+0.1i$     \\
\specialrule{0em}{1pt}{1pt}
\bottomrule[1.5pt]
\end{tabular}
\end{table}

\subsection{Isovector molecular states}

\begin{figure}[h!]
   \centering
   \includegraphics[scale=0.65,bb=395 251 17 495]{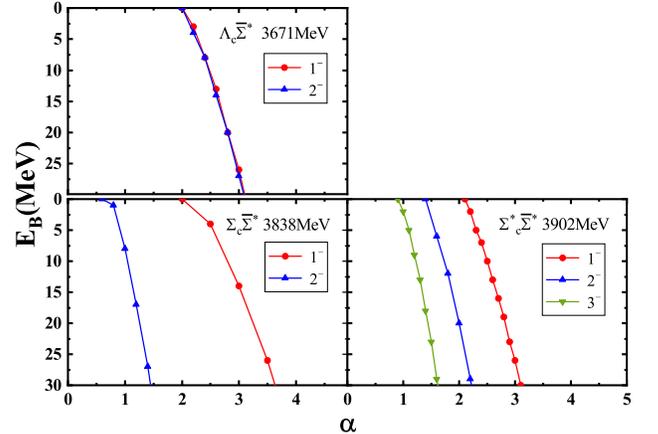}
   \caption{Binding energies of isovector bound states from the charmed-antistrange interactions  with the variation of $\alpha$ in single-channel calculation.}
   \label{1b}
 \end{figure}

In Fig.~\ref{1b}, the binding energies of all bound states produced from
isovector charmed-antistrange interactions are presented. The single-channel
calculation suggests that only seven bound states are produced from eighteen
channels considered. The $\Lambda_c\bar{\Sigma}^*$ states with $(1,2)^-$ are
almost degenerate, which all appear at $\alpha$ values of about 2.0, and its
binding energy increase to 30~MeV at  $\alpha$ values of about 3. The bound
states from the $\Sigma_c\bar{\Sigma}^*$ interaction with $(1,2)^-$ appears at
$\alpha$ values of about 0.5 and 2, respectively.  The three bound state from
the $\Sigma^*_c\bar{\Sigma}^*$ interaction with $(1,2,3)^-$ appears at $\alpha$
values about 1, 1.5, 2, respectively. These states are well distinguished, which
may make their  coexistence less possible.

 The coupled-channel results are presented in Table~\ref{t1b}. In the
 isovector case, the widths of the $\Sigma^*_c\bar{\Sigma}^*$ state with $(1,2,3)^-$ increase very rapidly to about $7$ MeV or larger with the increase of
 the parameter $\alpha$. The state with $1^-$ has strong coupling to the channels
 $\Sigma_c\bar{\Sigma}^*$, $\Sigma^*_c\bar{\Lambda}$, and
 $\Lambda_c\bar{\Sigma}$. The state with $2^-$ couples to channels
 $\Sigma_c\bar{\Sigma}^*$, $\Sigma^*_c\bar{\Sigma}$, and
 $\Sigma^*_c\bar{\Lambda}$. For the states with $3^-$, except the channels
 $\Sigma_c\bar{\Lambda}$ and $\Lambda_c\bar{\Sigma}$, other channels
 have strong couplings.  For the two  $\Lambda_c\bar{\Sigma}^*$ states with
 $(1,2)^-$, the $\Lambda_c\bar{\Sigma}$ channel is the dominant
 channel to produce their total widths.

\renewcommand\tabcolsep{0.41cm}
\renewcommand{\arraystretch}{1.22}
\begin{table*}[ht!]   
\footnotesize
\caption{The masses and widths of isovector charmed-antistrange molecular states at different values of $\alpha$. Other notations are the same as Table~\ref{t0a}.\label{t1b}}
\begin{tabular}{crrrrrrrrrrrrrrrr}\bottomrule[1.5pt]
\specialrule{0em}{1pt}{1pt}
$I=1$&$\alpha$ &\multicolumn{1}{c}{$CC$}& \multicolumn{1}{c}{$\Sigma_c\bar{\Sigma}^*$} & \multicolumn{1}{c}{$\Sigma^*_c\bar{\Sigma}$} & \multicolumn{1}{c}{$\Lambda_c\bar{\Sigma}^*$} &   \multicolumn{1}{c}{$\Sigma_c\bar{\Sigma}$} &\multicolumn{1}{c}{$\Sigma^*_c\bar{\Lambda}$}&\multicolumn{1}{c}{$\Sigma_c\bar{\Lambda}$} &\multicolumn{1}{c}{  $\Lambda_c\bar{\Sigma}$} \\
\specialrule{0em}{1pt}{1pt}
\hline
\specialrule{0em}{1pt}{1pt}
$\Sigma^*_c\bar{\Sigma}^*(1^-)$
&$2.2$ &$3+10.9i$   &$4+6.6i$     &$0+0.3i$      &$0+0.6i$         &$0+0.1i$   &$3+1.5i$      &$0+0.2i$  &$0+0.8i$    \\
3902 {\rm MeV}
&$2.4$ &$6+9.7i$   &$11+6.7i$     &$7+0.5i$      &$0+0.6i$         &$7+0.6i$   &$8+2.4i$      &$7+0.3i$  &$1+1.6i$    \\
&$2.7$ &$13+11.7i$   &$21+5.4i$    &$17+0.8i$      &$2+1.1i$  &$16+0.1i$   &$18+3.7i$      &$16+0.5i$  &$7+3.8i$   \\
&$2.8$ &$16+14.1i$   &$25+5.9i$    &$20+0.9i$       &$3+1.4i$  &$19+0.1i$   &$22+4.5i$      &$19+0.6i$  &$9+4.9i$   \\
\specialrule{0em}{1pt}{1pt}
$\Sigma^*_c\bar{\Sigma}^*(2^-)$
&$1.7$ &$0+5.1i$   &$0+2.3i$      &$0+1.6i$  &$0+1.0i$ &$1+0.0i$         &$0+1.6i$   &$1+0.1i$  &$0+0.2i$    \\
3902 {\rm MeV}
&$2.2$ &$13+9.8i$   &$0+7.2i$      &$13+1.0i$  &$0+1.5i$ &$15+0.1i$         &$10+4.3i$   &$15+0.4i$  &$10+1.2i$    \\
&$2.4$ &$17+11.3i$   &$2+9.1i$      &$20+6.3i$  &$2+1.5i$ &$23+0.2i$        &$17+5.5i$  &$23+0.6i$  &$18+2.0i$    \\
&$2.5$ &$19+12.9i$  &$4+10.5i$        &$24+7.1i$ &$3+1.4i$    &$27+0.2i$             &$21+6.1i$ &$27+0.7i$  &$21+2.4i$      \\
\specialrule{0em}{1pt}{1pt}
$\Sigma^*_c\bar{\Sigma}^*(3^-)$
&$1.0$ &$0+3.3i$   &$1+1.8i$    &$1+1.2i$      &$0+1.0i$  &$1+0.0i$  &$0+1.4i$      &$1+0.0i$  &$1+0.1i$     \\
3902 {\rm MeV}
&$1.1$ &$0+5.1i$   &$4+2.4i$    &$4+1.8i$      &$3+1.6i$  &$3+1.6i$  &$3+2.1i$      &$5+0.0i$  &$5+0.1i$     \\
&$1.2$ &$3+7.1i$   &$8+3.6i$    &$8+2.4i$       &$6+2.3i$  &$6+2.3i$ &$7+2.9i$      &$9+0.0i$  &$9+0.1i$      \\
\specialrule{0em}{1pt}{1pt}
$\Lambda_c\bar{\Sigma}^*(1^-)$
&$2.2$ &$6+3.0i$   &$--$     &$--$      &$--$  &$0+0.4i$ &$4+0.3i$  &$4+0.0i$     &$3+1.2i$      \\
3671 {\rm MeV}
&$2.4$ &$12+3.7i$   &$--$    &$--$      &$--$  &$9+0.3i$ &$8+0.3i$  &$10+0.0i$     &$8+1.8i$     \\
&$2.6$ &$18+4.1i$   &$--$    &$--$       &$--$  &$15+0.2i$ &$14+0.2i$   &$16+0.0i$     &$14+2.6i$    \\
\specialrule{0em}{1pt}{1pt}
$\Lambda_c\bar{\Sigma}^*(2^-)$
&$2.1$ &$4+3.3i$   &$--$     &$--$      &$--$  &$4+0.0i$ &$4+1.0i$  &$5+0.5i$     &$4+1.2i$      \\
3671 {\rm MeV}
&$2.4$ &$7+3.6i$   &$--$    &$--$      &$--$  &$9+0.0i$ &$11+0.9i$  &$10+0.6i$     &$9+1.8i$     \\
&$2.6$ &$12+4.0i$   &$--$    &$--$       &$--$  &$15+0.0i$ &$17+0.7i$   &$16+0.6i$     &$14+2.5i$    \\
\specialrule{0em}{1pt}{1pt}
\bottomrule[1.5pt]
\end{tabular}
\end{table*}

\subsection{Isotensor molecular states}

\begin{figure}[h!]
   \centering
   \includegraphics[scale=0.65,bb=395 231 17 495]{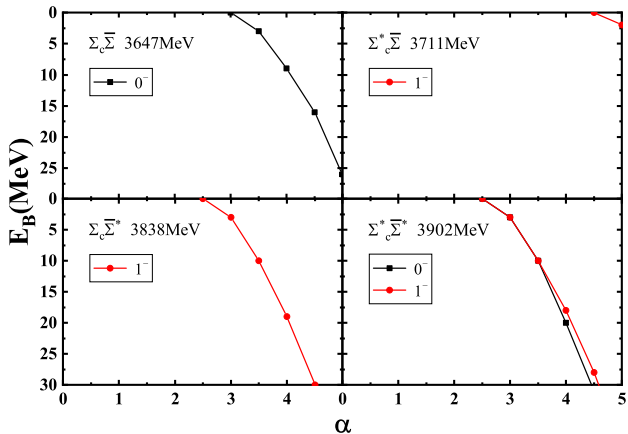}
   \caption{Binding energies of isotensor bound states from the charmed-antistrange interactions  with the variation of $\alpha$ in single-channel calculation.}
   \label{2b}
 \end{figure}

In Fig.~\ref{2b}, the binding energies of all bound states produced from
isotensor charmed-antistrange interactions are presented. There are only five
bound states can be produced from ten channels considered within the reasonable
range of cutoffs. The $\Sigma_c\bar{\Sigma}$ state  with $0^-$ can be produced
at an $\alpha$ value of about $3.0$.  The $\Sigma_c\bar{\Sigma}^*$ state with
$1^-$ and the $\Sigma^*_c\bar{\Sigma}^*$ states with $(0,1)^-$ can be produced at
an $\alpha$ value of about 2.5, while the $\Sigma_c\bar{\Sigma}^*$ with $1^-$
appears at an $\alpha$ value of about 4.5.  Generally speaking, the $\alpha$
value to produce  these states are considerable larger than these to produce
most of the bound states in the above, which suggest weak  attraction in these
channels. The possibility of existence of such states is also very low.

The coupled channel results are presented in Table~\ref{t2b}. It can be seen
that large $\alpha$ values are still required to produce the poles.  For the two
$\Sigma^*_c\bar{\Sigma}\textcolor[rgb]{1.00,0.00,0.00}{^*}$ states with $(0,1)^-$, the dominant channel is the
$\Sigma_c\bar{\Sigma}^*$ channel. The $\Sigma_c\bar{\Sigma}^*$ state with $1^-$
strongly couples to the $\Sigma_c^*\bar{\Sigma}$ channel. Though there exists
only one channel $\Sigma_c\bar{\Sigma}$ below the  $\Sigma_c^*\bar{\Sigma}$
threshold, the $\Sigma_c^*\bar{\Sigma}$  state with $1^-$ acquire considerable
large width.

 \renewcommand\tabcolsep{0.22cm}
 \renewcommand{\arraystretch}{1.25}
 \begin{table} [h!]   
 \footnotesize
 \caption{The masses and widths of isotensor charmed-antistrange molecular states at different values of $\alpha$. Other notations are the same as Table~\ref{t0a}.\label{t2b}}
 \begin{tabular}{crrrrrrrr}\bottomrule[1.5pt]
\specialrule{0em}{1pt}{1pt}
 $I=2$&$\alpha$ &\multicolumn{1}{c}{$CC$}& \multicolumn{1}{c}{$\Sigma_c\bar{\Sigma}^*$} & \multicolumn{1}{c}{$\Sigma^*_c\bar{\Sigma}$} &   \multicolumn{1}{c}{$\Sigma_c\bar{\Sigma}$}  \\
\specialrule{0em}{1pt}{1pt}
 \hline
\specialrule{0em}{1pt}{1pt}
$\Sigma^*_c\bar{\Sigma}^*(0^-)$
 &$3.0$ &$2+2.1i$   &$2+2.1i$     &$2+0.2i$      &$3+0.0i$       \\
 3902 {\rm MeV}
 &$3.5$ &$8+4.8i$   &$10+4.6i$    &$9+0.9i$      &$10+0.0i$      \\
 &$4.0$ &$16+12.0i$   &$19+10.5i$    &$17+2.4i$       &$15+0.0i$      \\
\specialrule{0em}{1pt}{1pt}
 $\Sigma^*_c\bar{\Sigma}^*(1^-)$
 &$3.0$ &$0+1.7i$   &$0+1.8i$     &$0+0.1i$      &$0+0.0i$        \\
 3902 {\rm MeV}
 &$3.5$ &$2+4.3i$   &$3+4.2i$    &$3+0.3i$      &$3+0.1i$      \\
 &$4.0$ &$9+7.2i$   &$10+6.3i$    &$9+0.7i$       &$9+0.3i$     \\
\specialrule{0em}{1pt}{1pt}
$\Sigma_c\bar{\Sigma}^*(1^-)$
 &$3.0$ &$2+0.6i$   &$--$          &$3+0.4i$  &$3+0.1i$      \\
 3838 {\rm MeV}
 &$3.5$ &$8+1.8i$   &$--$         &$10+1.2i$  &$9+0.3i$     \\
 &$4.0$ &$15+4.3i$   &$--$    &$19+2.2i$       &$16+1.0i$     \\
\specialrule{0em}{1pt}{1pt}
$\Sigma^*_c\bar{\Sigma}(1^-)$
 &$4.5$ &$0+7.7i$   &$--$     &$--$      &$0+7.7i$       \\
 3711 {\rm MeV}
 &$4.9$ &$1+10.6i$   &$--$    &$--$       &$1+10.6i$      \\
 &$5.0$ &$2+10.5i$   &$--$    &$--$      &$2+10.5i$       \\
\specialrule{0em}{1pt}{1pt}
 \bottomrule[1.5pt]
 \end{tabular}
 \end{table}

\section{Summary and discussion}\label{Sec: summary}

In this work, we systematically study the charmed-strange molecular states with
quantum numbers $C$=1 and S=$\pm 1$, in a qBSE approach together with the
one-boson-exchange model. The potential kernels are constructed with the help of
the effective Lagrangians with $SU(3)$, chiral and heavy quark symmetries. With
the exchange potential obtained, the S-wave bound states are searched for as the
pole of the scattering amplitudes.

The attractions widely exist in the charmed-strange interactions. In the current
work, we consider all $S$-wave interactions of a charmed baryon and a strange
baryon $\Lambda_c\Lambda$, $\Lambda_c\Sigma^{(*)}$, $\Sigma_c^{(*)}\Lambda$, and
$\Sigma^{(*)}_c\Sigma^{(*)}$, which results in 40 channels with different spin
parities. Among these channels, 39 bound states are produced in the range of the
parameter considered in the current work. Among these bound states, 32 states
can be produced from the interactions in a range of $\alpha$ value from 0 to 1.
For channels that have two or three different isospins, the bound state with
smaller isospin, the binding energies tend to larger than the state with
larger isospin. The coupled-channel calculations do not change the conclusion
from the single-channel calculations, and the possible strong couplings are also
suggested.

Compared with the charmed-strange interactions, fewer states can be produced in
the charmed-antistrange interactions. There are also  40 $S$-wave channels as in
the charmed-strange sector. However,  only 17 states can be produced. Besides,
the attractions of many states are weak as suggested by the $\alpha$ values to
produce these states. Among 17 states produced, only 4 states can be produced
with an $\alpha$ value smaller than 1. For the states from  charmed-antistrange
interactions, the quark and antiquark in the
systems can be annihilated, which results in quantum numbers as a charm-strange
meson $D_s$.  It also makes it easy to be found in experiment.  Hence, we
suggest the experimental research for such states

\vskip 10pt

\noindent {\bf Acknowledgement} This project is supported by the Postgraduate
Research and Practice Innovation Program of Jiangsu Province (Grants No.
KYCX22\_1541) and the National Natural Science Foundation of China (Grants No.
11675228).

\end{document}